\documentstyle[emulateapj,apjfonts]{article}

\def\arcsec{\ifmmode '' \else $''$\fi}
\def\arcmin{\ifmmode ' \else $'$\fi}
\def\arcsecpoint{\ifmmode ''\!. \else $''\!.$\fi}
\def\arcminpoint{\ifmmode '\!. \else $'\!.$\fi}
\def\cc{\ifmmode {\rm cm}^{-3} \else cm$^{-3}$\fi}
\def\cl{\ifmmode {\rm cm}^{-2} \else cm$^{-2}$\fi}
\def\micron{\ifmmode \mu{\rm m} \else $\mu$m\fi}
\def\kms{\ifmmode {\rm km\,s}^{-1} \else km\,s$^{-1}$\fi}
\def\Hubble{\ifmmode {\rm km\,s}^{-1}\,{\rm Mpc}^{-1}
        \else km\,s$^{-1}$\,Mpc$^{-1}$\fi}
\def\ergsec{\ifmmode {\rm ergs\;s}^{-1} \else ergs s$^{-1}$\fi}
\def\ergscm{\ifmmode {\rm ergs\,s}^{-1}\,{\rm cm}^{-2}
          \else ergs\,s$^{-1}$\,cm$^{-2}$\fi}
\def\ergscmA{\ifmmode {\rm ergs\,s}^{-1}\,{\rm cm}^{-2}\,{\rm \AA}^{-1}
          \else ergs\,s$^{-1}$\,cm$^{-2}$\,\AA$^{-1}$\fi}
\def\ergscmHz{\ifmmode {\rm ergs\,s}^{-1}\,{\rm cm}^{-2}\,{\rm Hz}^{-1}
          \else ergs\,s$^{-1}$\,cm$^{-2}$\,Hz$^{-1}$\fi}
\def\Msun{\ifmmode M_{\odot} \else $M_{\odot}$\fi}
\def\Lsun{\ifmmode L_{\odot} \else $L_{\odot}$\fi}
\def\qo{\ifmmode q_{0} \else $q_{0}$\fi}
\def\Ho{\ifmmode H_{0} \else $H_{0}$\fi}
\def\ciii{C\,{\sc iii}}

\newcommand{\ovi}{O~{\sc vi}}
\newcommand{\ovii}{O~{\sc vii}}
\newcommand{\oviii}{O~{\sc viii}}
\newcommand{\heii}{He~{\sc ii}}
\newcommand{\feii}{Fe~{\sc ii}}

\lefthead{KRISS ET AL.}
\righthead{FUSE SPECTRUM OF MRK 509}

\begin{document}
\submitted{To appear in ApJ Letters}

\title{FUSE Observations of Intrinsic Absorption
in the Seyfert 1 Galaxy Mrk~509}

\author{
G. A. Kriss\altaffilmark{1,3},
R. F. Green\altaffilmark{2},
M. Brotherton\altaffilmark{2},
W. Oegerle\altaffilmark{3},
K. R. Sembach\altaffilmark{3},
A. F. Davidsen\altaffilmark{3},
S. D. Friedman\altaffilmark{3},
M. E. Kaiser\altaffilmark{3},
W. Zheng\altaffilmark{3},
B. Woodgate\altaffilmark{4},
J. Hutchings\altaffilmark{5},
J. M. Shull\altaffilmark{6}
D. G. York\altaffilmark{7}
}
\altaffiltext{1}{Space Telescope Science Institute,
	3700 San Martin Drive, Baltimore, MD 21218; gak@stsci.edu}
\altaffiltext{2}{Kitt Peak National Observatory,
	National Optical Astronomy Observatories, P.O. Box 26732,
	950 North Cherry Ave., Tucson, AZ, 85726-6732}
\altaffiltext{3}{Center for Astrophysical Sciences, Department of Physics and
	Astronomy, The Johns Hopkins University, Baltimore, MD 21218--2686}
\altaffiltext{4}{Laboratory for Astronomy and Solar Physics, Code 681,
	NASA/Goddard Space Flight Center, Greenbelt, MD 20771}
\altaffiltext{5}{Dominion Astrophysical Observatory, National Research Council
	of Canada, Victoria, BC, V8X 4M6, Canada; john.hutchings@hia.nrc.ca}
\altaffiltext{6}{CASA and JILA, Department of
	Astrophysical and Planetary Sciences, University of Colorado,
	Campus Box 389, Boulder, CO 80309; mshull@casa.colorado.edu}
\altaffiltext{7}{Department of Astoronomy,
	University of Chicago, Chicago, IL 60637; don@oddjob.uchicago.edu}

\begin{abstract}
We present far-ultraviolet spectra of the Seyfert 1 galaxy Mrk~509 obtained in
1999 November with the Far Ultraviolet Spectroscopic Explorer (FUSE).
Our data span the observed wavelength range 915--1185 \AA\ at a resolution
of $\sim20~\kms$.  The spectrum shows a blue continuum, broad
\ovi\ $\lambda\lambda1032,1038$ emission, and a
broad \ciii\ $\lambda977$ emission line.
Superposed on these emission components,
we resolve associated absorption lines of \ovi\ $\lambda\lambda1032,1038$,
\ciii\ $\lambda977$, and Lyman lines through L$\zeta$.  Seven distinct
kinematic components are present, spanning a velocity range of
$-440~\rm to +170~\kms$ relative to the systemic velocity.
The absorption is clustered in two groups, one centered at $-370~\rm \kms$
and another at the systemic velocity.  The blue-shifted cluster may be
associated with the extended line emission visible in deep images of Mrk~509
obtained by Phillips et al.  Although several components appear to be saturated,
they are not black at their centers.
Partial covering or scattering permits $\sim 7$\% of the broad-line or
continuum flux to be unaffected by absorption.
Of the multiple components, only one has
the same ionization state and column density as highly ionized gas that
produces the \ovii\ and \oviii\ ionization edges in X-ray spectra of Mrk~509.
\end{abstract}


\section{INTRODUCTION}

Mrk 509 straddles the boundary in luminosity between Seyfert 1 nuclei and QSOs
with an absolute magnitude of $M_B = -22.0$.  This gives it particular
importance for understanding how the properties typical of the well-studied
nearby active galactic nuclei (AGN) might scale with luminosity, and, therefore,
be applied to the more distant and luminous QSOs.
Like many Seyfert 1 galaxies, Mrk~509 exhibits intrinsic UV absorption
lines (\cite{York84}; \cite{Crenshaw95}; \cite{Savage97}; \cite{Crenshaw99})
and X-ray absorption edges of \ovii\ and \oviii\ (\cite{Reynolds97};
\cite{George98}).
The UV absorption is blue-shifted relative to the systemic velocity of
10,365 \kms\ (\cite{Phillips83}), and Phillips et al.\markcite{Phillips83}
suggest that it is related to the expanding shell of high ionization gas
visible in images and spectra out to a radius of 15\arcsec\ from the nucleus.

\begin{figure*}[t]
\plotfiddle{"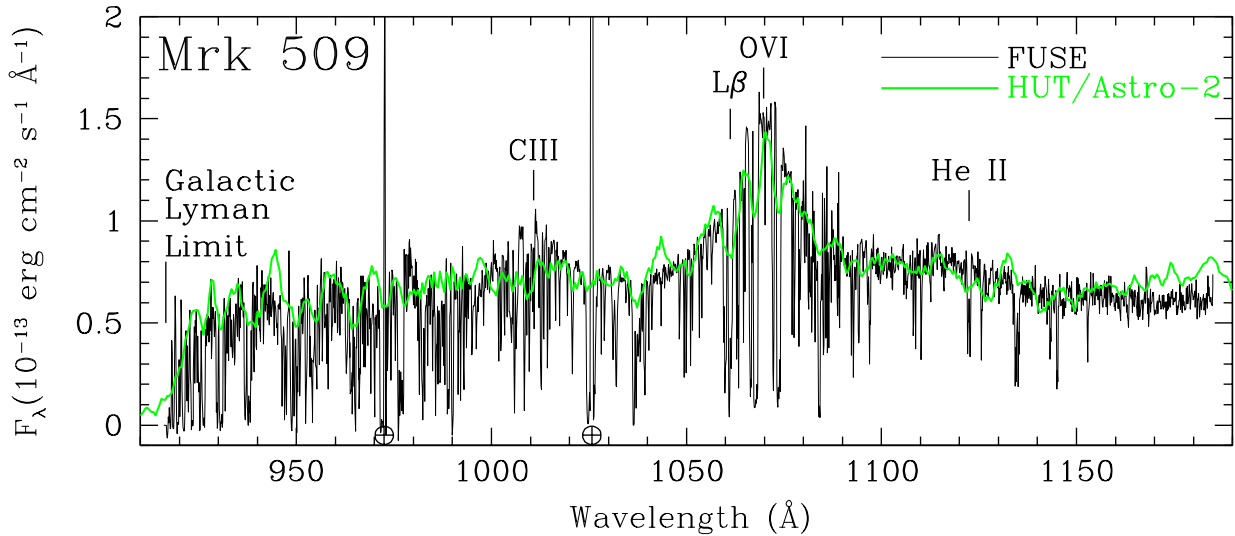"}{2.55 in}{-90}{77}{77}{-300}{435}
\caption{
FUSE spectrum of Mrk 509 (binned by 20 pixels) is shown as a thin black line.
Hopkins Ultraviolet Telescope (HUT) data obtained during
the Astro-2 mission in 1995 March (scaled by a factor of 0.69)
are shown as the green curve.
\vskip -14pt
\label{fig1.ps}}
\end{figure*}

At high resolution, UV absorption in Seyferts appears to be kinematically
complex (NGC~3516: \cite{Crenshaw98}; NGC~5548: \cite{Mathur99}), with
at most one of the UV components having any possible relation to the X-ray
absorbing gas (\cite{Mathur99}).  High resolution observations
of UV absorption lines covering a range of ionization states is the key
to determining which, if any, of the UV absorbers are associated with
an X-ray warm absorber.
The FUSE observations of Mrk~509 we present in this paper extend the far-UV
spectral coverage to wavelengths shortward of 1200 \AA.
Our spectra include the \ovi\ $\lambda\lambda1032,1038$ resonance doublet
and the high-order Lyman lines, down to the redshifted Lyman limit.
Since the L$\alpha$ lines resolved in earlier UV observations are saturated,
the high-order lines visible with FUSE provide a better constraint on the
total neutral hydrogen column density.
The \ovi\ doublet is a crucial link for establishing a connection between
the higher ionization absorption edges seen in the X-ray and the lower
ionization absorption lines seen in earlier UV observations.
Based on these new observations, we discuss the implications for the location
of the absorbing gas in AGN, and how intrinsic absorption in low-$z$ AGN
may be related to the broad-absorption line phenomenon in more luminous QSOs.

\section{OBSERVATIONS}

FUSE comprises four separate primary mirrors gathering light for
four prime-focus, Rowland-circle spectrographs and
two, two-dimensional, photon-counting detectors.
For a full description of FUSE, its mission, and its in-flight performance,
see Moos et al. (2000)\markcite{Moos2000},
and Sahnow et al. (2000)\markcite{Sahnow2000}.
Two of the optical systems employ LiF coatings on the optics, giving
coverage from $\sim990$--1187 \AA, and the other two use SiC coatings,
which provide reflectivity down to wavelengths as short as 905 \AA.
The mirror systems focus light on a slit assembly in the focal plane.
The holographically ruled gratings disperse the light entering through the
slits and form an astigmatic image on the two-dimensional
microchannel-plate detectors.
The detectors have KBr photocathodes and delay-line anode readouts
that provide the location and arrival time of each photon event.

Mrk 509 was observed on 1999 Nov 9 and 1999 Nov 11 through the
$30\arcsec \times 30\arcsec$ low-resolution apertures.
We obtained good spectra from the LiF1 and LiF2 channels covering the
987--1187 \AA\ band, and lower signal-to-noise-ratio (S/N) spectra from the
SiC2 channel covering 905--1091 \AA.
As a result of channel misalignment during the observations,
almost no data were obtained in the SiC1 channel.
We recorded the data in photon-address mode, which provides a time-tagged list
of event positions in the down-linked data stream.
This enables us to filter out short periods of event ``bursts"
(\cite{Sahnow2000}) and to correct for image motion on the detectors
to achieve the best spectral resolution.
As described by Sahnow et al. (2000)\markcite{Sahnow2000}, we extracted
one-dimensional spectra from the two-dimensional data on each of the
active detector segments.
These extracted spectra are dark subtracted and flux and wavelength calibrated
by the standard FUSE calibration pipeline.
We estimate that the flux scale is accurate to $\sim$10\%, and that wavelengths
are accurate to $\sim$15 \kms.
Poisson errors and data quality flags are propagated through the data reduction
process along with the science data.

To produce the full spectrum shown in Figure \ref{fig1.ps},
we have spliced together sections of data from segments SiC2A, LiF1A, SiC2B,
LiF1B, and LiF2A to eliminate the gaps in wavelength coverage and to present
the best S/N.  These data have also been binned
by 20 pixels (0.12 \AA) to show the overall appearance of
the spectrum.  In our analysis discussed in the following sections, we use
spectra binned by only 5 pixels.  This preserves the full spectral resolution
of $\sim 20 ~\kms$ for this observation.

In Figure \ref{fig1.ps} strong, broad \ovi\ emission is readily apparent as is
broad \ciii\ $\lambda$977 emission.
Another less prominent feature in the spectrum is the hump of emission
redward of \ovi\ $\lambda\lambda1032,1038$.
We have marked the position of \heii\ $\lambda1085$, but one can see that
its wavelength is a bit too long.
A similar feature was noted in the spectra of low-redshift quasars by
Laor et al. (1995)\markcite{Laor95} who suggested this may be blended
\feii\ emission.
The numerous Galactic absorption features, particularly $\rm H_2$,
make the intrinsic spectrum difficult to see shortward of 1000 \AA, but, at the
resolution of FUSE, one can trace out the intrinsic spectrum as the peaks
between the foreground absorptions.
Note how the \ciii\ $\lambda$977 line is not prominent in the overlayed
Hopkins Ultraviolet Telescope (HUT) data, primarily because the
foreground Galactic absorption renders it
less visible at the $\sim3$\AA\ resolution of HUT.

\section{WARM ABSORBING GAS IN MRK 509}

A detailed examination of the FUSE spectrum at full spectral resolution shows
that absorption near the redshift of Mrk~509 is visible 
in the Lyman lines, in the \ovi\ $\lambda\lambda$1032,1038 resonance doublet,
and in \ciii\ $\lambda$977.
Figure 2 shows portions of the Mrk~509 spectrum around each
absorption line complex with velocities relative to the AGN systemic velocity.

Close inspection of the \ovi\ doublet and the Lyman lines shows that while
much of the absorption appears to be saturated, the absorption troughs are
not black. (Scattered light in FUSE at these levels is negligible.  See
\cite{Sahnow2000}.)
Partial covering by the absorbers or scattering around the absorbing region
can explain this appearance.
Extended broad-line Balmer emission and variable, polarized broad Balmer-line
emission are seen in Mrk~509 (\cite{Mediavilla98}; \cite{TM88}; \cite{Young99}).
These are both indications that scattering may play some role
in producing the light at the bottoms of the absorption troughs.

In the L$\beta$ and \ovi\ absorption lines we have identified at least 7
distinct kinematic components spanning a range of velocities from
$-478$ to $+166$ \kms\ relative to the systemic velocity of Mrk~509.
To ascertain the physical properties of these intrinsic absorbers in Mrk 509,
we fit a model to the spectral regions surrounding the \ovi\ lines, the Lyman
lines, and \ciii\ $\lambda$977
using the IRAF task {\tt specfit} (\cite{Kriss94}).
Figure 3 shows the best fit overlayed on the L$\beta$/\ovi\ region.
Our model of the emission components includes an underlying power-law continuum,
a pair of broad (FWHM$\sim$11,000 \kms) \ovi\ emission lines with their
relative intensities fixed at the optically thin ratio of 2:1, a pair of narrow
(FWHM$\sim$1200 \kms) \ovi\ emission lines, again with their relative
intensities fixed at a 2:1 ratio, and a broad L$\beta$ emission line with its
width and velocity linked to those of the broad \ovi\ lines.
The absorption lines are treated as Gaussians in optical depth, and they are
allowed to partially cover the emission components with fraction $f_c$.
The wavelengths of the \ovi\ doublets are linked at the ratio of their
laboratory values, their velocity widths are required to be identical,
and their relative optical depths are fixed at a 2:1 ratio.
Thus, a pair of lines uniquely determines the column density and
the covering fraction for a given kinematic component.
The L$\beta$ lines have their wavelengths linked to those of the \ovi\ lines,
but, to allow for residual uncertainties in the FUSE wavelength scale,
we permit a linear adjustment to the whole group of L$\beta$ lines.
The widths and the covering fractions for the L$\beta$ lines are fixed at the
values determined for the \ovi\ lines.

\vskip -10pt
\vbox to 4.30in {
\vbox to 14pt{\vfill}
\plotfiddle{"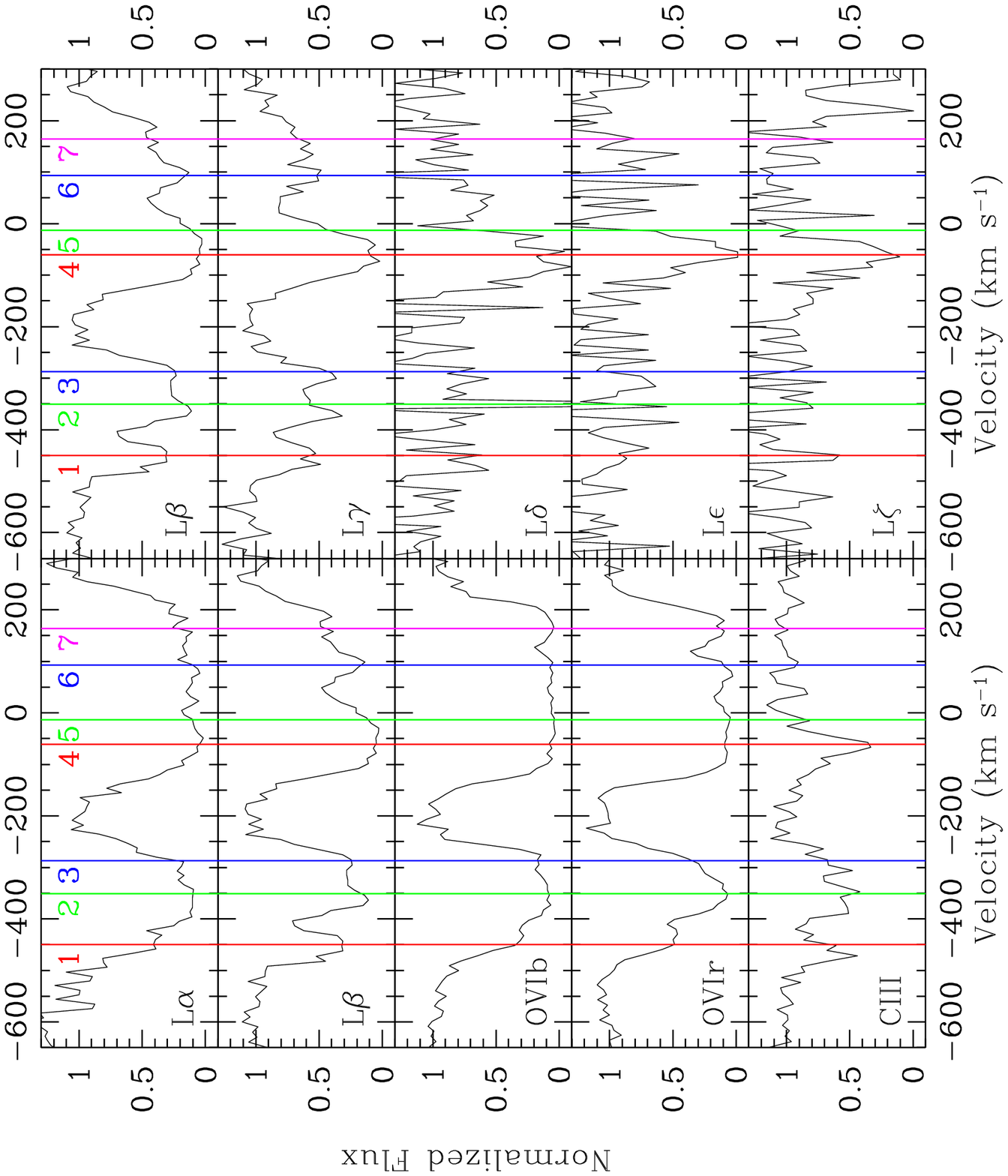"}{3.00 in}{-90}{44}{44}{-178}{250}
\parbox{3.5in}{
\small\baselineskip 9pt
\footnotesize
\indent
{\sc Fig.}~2.---
Normalized line profiles for the Lyman lines, {\sc C iii} $\lambda 977$,
{\sc O VI} $\lambda 1032$ (OVIb), and {\sc O VI} $\lambda 1038$ (OVIr) are
shown.  These data are binned by 5 pixels, and foreground Galactic
absorption features have been divided out.  The central wavelengths of
the 7 distinct intrinsic absorption components are marked.
The L$\alpha$ profile at the upper left is archival HST data
from a GHRS observation by B. Savage.
\label{fig2.ps}}
\vbox to 14pt{\vfill}
}
\vskip -10pt

Higher-order Lyman lines are visible in the FUSE spectrum out to L$\zeta$.
These lines are also fit simultaneously with L$\beta$.  Their optical
depths are fixed at the ratio of their oscillator strengths to that of L$\beta$,
their wavelengths are linked to those of \ovi\ at the ratio of the
vacuum wavelengths (again allowing for slight linear adjustments), and
their widths are fixed at the same values as those of the \ovi\ components.
The Lyman lines permit an independent check of the covering fractions determined
from the \ovi\ doublets.  We find that if they are allowed to vary
independently, they give results consistent with those determined
from \ovi\ alone, so we leave them fixed at the \ovi\ values in our
final results.
\ciii\ $\lambda 977$ is treated similarly.  Although there is no independent
check on its covering fraction, we also fixed $f_c$ at the \ovi\ values.

The results of our fits for \ciii\ $\lambda 977$, L$\beta$, and
\ovi\ $\lambda 1032$ are given in Table 1.
The column densities for each ion are determined by integrating the optical
depth across each parameterized line profile.
(Since the parameters
given completely determine the values for \ovi\ $\lambda 1038$ and the
remaining Lyman lines, these are not shown.)

To determine physical conditions in the absorption components, we used
photoionization models similar to those used by
Krolik \& Kriss (1995)\markcite{KK95} and
Kriss et al. (1996).\markcite{Kriss96a} From a grid of models we determined
the total column density and the ionization parameter based on the observed
relative columns of {\sc H~i} and \ovi.
With no other constraints, this method can lead to double-valued results for
the ionization parameter and column density since the ratio of \ovi\ to
{\sc H~i} will rise to a peak and then decline.  However, we note that the
presence of \ciii\ in components 1--4 restricts the solutions to the lower
ionization parameter in these cases.  Component 5 lies near the peak of the
N(\ovi)/N({\sc H~i}) curve, and, as we show below, the inferred ionization and
column density are corroborated by X-ray observations of \ovii\ and \oviii.
As component 5 fully accounts for the observed X-ray absorption, a high-column,
high-ionization-parameter solution for components 6 and 7 is also ruled out
by the X-ray observations.
Physical parameters for the 7 kinematic components are shown in Table 2.
Note that most components have relatively low total column densities and
ionization parameters.  The components associated with the most blue-shifted
complex could well be associated with the outflowing narrow-emission line gas
as originally suggested by Phillips et al. (1983)\markcite{Phillips83}.
This would place it many kiloparsecs from the central ionizing source.
Similarly, the lower ionization components (6 \& 7) near the systemic velocity
may be associated with the low-ionization gas in the rotating disk near the
center of Mrk~509 (\cite{Phillips83}).

\vskip -10pt
\vbox to 4.00in {
\vbox to 14pt{\vfill}
\plotfiddle{"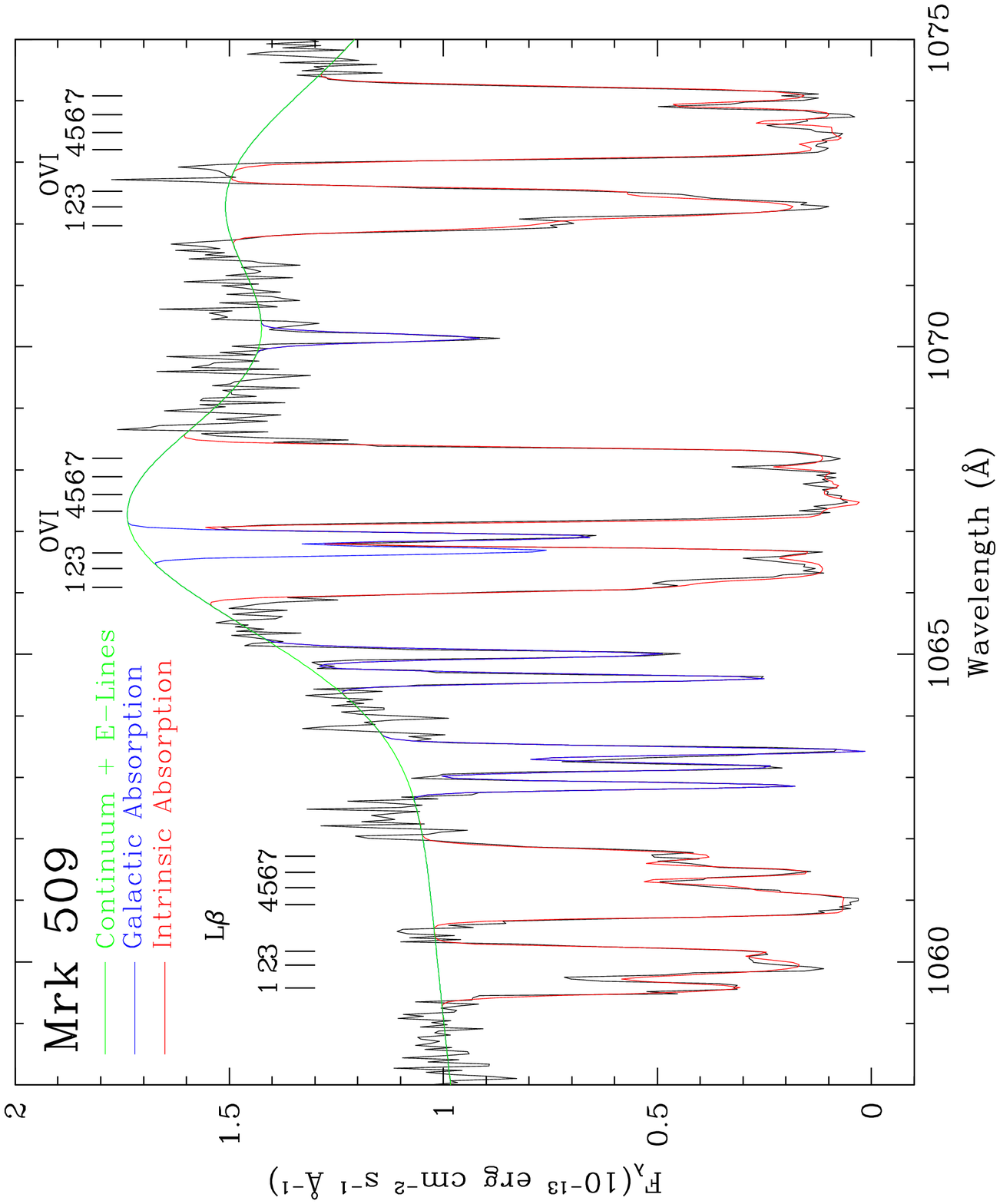"}{2.80 in}{-90}{38}{38}{-153}{230}
\parbox{3.5in}{
\small\baselineskip 9pt
\footnotesize
\indent
{\sc Fig.}~3.---
FUSE spectrum of Mrk 509 (binned by 5 pixels) in the L$\beta$/\ovi\ region
is shown as the thin black line.  The best fit described in the text is
overlayed in color. The thin green line shows the fitted continuum and emission
components. The thin red line shows the fitted intrinsic absorption
components, and the thin blue line shows the foreground Galactic absorption
lines. The central wavelengths of the 7 distinct intrinsic absorption
components are marked.
\label{fig3.ps}}
\vbox to 14pt{\vfill}
}
\vskip -4pt

Among all the absorption components, the most exceptional is \#5.
Its high ionization parameter and high total column density make it
likely to be directly associated with the intrinsic X-ray warm absorber.
In his analysis of the ASCA X-ray spectrum of Mrk~509,
Reynolds (1997)\markcite{Reynolds97} found \ovii\ and \oviii\ optical
depths of $0.11^{+0.03}_{-0.04}$ and $0.04^{+0.04}_{-0.03}$, respectively.
For threshold photoionization cross sections of
$\sigma_{O7} = 0.239 \times 10^{-18}~\rm cm^2$ and
$\sigma_{O8} = 0.109 \times 10^{-18}~\rm cm^2$ (\cite{RM79}), this implies
column densities of
$N_{O7} = (4.6^{+1.3}_{-1.7}) \times 10^{17}~\rm cm^{-2}$ and
$N_{O8} = (3.7^{+3.7}_{-2.8}) \times 10^{17}~\rm cm^{-2}$.
Our photoionization modeling for UV absorption component 5 predicts
$N_{O7} = 2.3 \times 10^{17}~\rm cm^{-2}$ and
$N_{O8} = 0.1 \times 10^{17}~\rm cm^{-2}$.
Considering the large uncertainties in the X-ray columns and the temporal
difference between the X-ray and UV observations,
the agreement is remarkably close.
The X-ray absorption predicted by our photoionization modeling for the other
UV absorption components is negligible.

We conclude that observations of the \ovi\ doublet in AGN is a perfect
complement to observations of the \ovii\ and \oviii\ 
edges often seen in the X-ray spectra of Seyfert 1 galaxies.
Comparison of the UV and X-ray measurements
permits us to rigorously test whether the gas responsible for
X-ray absorption in Seyferts also gives rise to the UV absorption,
as suggested by
Mathur et al. (1994\markcite{Mathur94}, 1995\markcite{Mathur95}).
The multiple kinematic
components, the wide range of ionization parameters, and the typical
low total columns inferred for most of the UV absorbers
(see \cite{Kriss96a}; \cite{Crenshaw99})
make it unlikely that the UV and X-ray absorption arise in the same gas.
In fact, in Mrk 509, with the high spectral resolution and far-UV
sensitivity of FUSE, we can easily pick out the high ionization
absorption component that is likely to be directly associated with
the X-ray absorbing gas.  The remaining components (which dominate
the total UV absorption) are lower column density and lower ionization.
This suggests that the absorbing medium is complex, with separate
UV and X-ray dominant zones.  One potential geometry is high density,
low column UV-absorbing clouds embedded in a low density,
high ionization medium that dominates the X-ray absorption.
This is possibly a wind driven off the obscuring torus or the accretion disk.

\vskip -10pt
\vbox to 4.0in {
\begin{center}
\small
{\sc TABLE 1\\
Absorption Lines in Mrk~509}
\vskip 2pt
\footnotesize
\begin{tabular}{lcccccc}
\hline
\hline
Feature & \# & $W_\lambda$ & {$\rm N_{ion}$} &
$\Delta \rm v ^a$ & FWHM & $f_c$ \\
  &  & (\AA) & $\rm (cm^{-2})$  & ($\rm km~s^{-1}$) & ($\rm km~s^{-1}$)  & \\
\hline
C III & 1 & 0.11 & $1.6\times10^{13}$ & $-438$ & 49 & 0.77 \\
$\lambda977.02$ & 2 & 0.21 & $2.4\times10^{13}$ & $-349$ & 63 & 0.91 \\
 & 3 & 0.06 & $5.4\times10^{12}$ & $-280$ & 31 & 1.00 \\
 & 4 & 0.17 & $2.1\times10^{13}$ & $-75$  & 57 & 0.92 \\
 & 5 & $<0.01$ & $<2.0\times10^{12}$ & $-5$ & 40 & 0.92 \\
 & 6 & $<0.01$ & $<2.0\times10^{12}$ & $+71$& 48 & 0.94 \\
 & 7 & $<0.01$ & $<2.0\times10^{12}$ & $+166$& 59& 0.94 \\
H I  & 1 & 0.25 & $5.7\times10^{14}$ & $-438$ & 49 & 0.77 \\
$\lambda1025.72$  & 2 & 0.24 & $5.3\times10^{14}$ & $-349$ & 63 & 0.91 \\
 & 3 & 0.26 & $9.3\times10^{14}$ & $-280$ & 31 & 1.00 \\
 & 4 & 0.43 & $6.0\times10^{15}$ & $-75$  & 57 & 0.92 \\
 & 5 & 0.36 & $2.7\times10^{14}$ & $-5$   & 40 & 0.92 \\
 & 6 & 0.25 & $1.2\times10^{15}$ & $+71$  & 48 & 0.94 \\
 & 7 & 0.28 & $1.2\times10^{15}$ & $+166$ & 59 & 0.94 \\
O VI & 1 & 0.24 & $1.9\times10^{14}$ & $-438$ & 49 & 0.77 \\
$\lambda1031.93$ & 2 & 0.46 & $1.3\times10^{15}$ & $-349$ & 63 & 0.91 \\
 & 3 & 0.14 & $1.2\times10^{14}$ & $-280$ & 31 & 1.00 \\
 & 4 & 0.44 & $1.2\times10^{15}$ & $-75$  & 57 & 0.92 \\
 & 5 & 0.31 & $3.2\times10^{15}$ & $-5$   & 40 & 0.92 \\
 & 6 & 0.37 & $1.2\times10^{15}$ & $+71$  & 48 & 0.94 \\
 & 7 & 0.41 & $9.4\times10^{14}$ & $+166$ & 59 & 0.94 \\
\hline
\end{tabular}
\vskip 2pt
\parbox{3.5in}{
\small\baselineskip 9pt
\footnotesize
\indent
$\rm ^a$Velocity relative to a systemic redshift of
$cz = 10365~\rm km~s^{-1}$ (Phillips et al. 1983\markcite{Phillips83}).}
\end{center}
}
\vskip -10pt

One puzzling inconsistency with the outflow hypothesis, however, is that
component 5 is at rest with respect to the systemic velocity.
If the X-ray absorbing gas is in the low-ionization disk observed by
Phillips et al. (1983)\markcite{Phillips83}, then the low line-of-sight velocity
could be explained by having most of its motion transverse to the line of
sight.  The ionization parameter $U \sim 0.4$ is consistent with this
location if the gas density is low enough--- for
$n = 10^3~\rm cm^{-3}$ and an ionizing luminosity of
$L_{ion} = 3.4 \times 10^{45}~\rm erg~s^{-1}$ for Mrk~509,
the absorbing gas would be located at a distance of 330 pc
from the nuclear source, or $\sim 0.5''$ for $H_0 = 65~\rm km~s^{-1}~Mpc^{-1}$.
Setting better constraints on the location of the X-ray absorbing gas, however,
will require better knowledge of the gas density, which can be determined
from variability studies
(e.g., \cite{Kriss97}; \cite{Hamann97}; \cite{Espey98}).

\vbox to 1.75in {
\begin{center}
\small
{\sc TABLE 2\\
Physical Properties of the Absorbers in Mrk~509}
\vskip 2pt
\footnotesize
\begin{tabular}{lccc}
\hline
\hline
\# & $\rm N_{OVI} / N_{HI}$ &{$\rm N_{tot}$} & log U \\
    &   &  $\rm ( cm^{-2} )$  &   \\
\hline
1 & \phantom{0}0.37 & $2.0\times10^{18}$ & $-1.64$ \nl
2 & \phantom{0}1.51 & $5.9\times10^{18}$ & $-1.19$ \nl
3 & \phantom{0}0.48 & $2.2\times10^{18}$ & $-1.79$ \nl
4 & \phantom{0}0.19 & $1.6\times10^{19}$ & $-1.73$ \nl
5 &           13.9  & $5.0\times10^{20}$ & $-0.43$ \nl
6 & \phantom{0}2.14 & $7.6\times10^{18}$ & $-1.41$ \nl
7 & \phantom{0}2.76 & $6.8\times10^{18}$ & $-1.46$ \nl
\hline
\end{tabular}
\end{center}
\vskip -15pt
}

\section{SUMMARY}

The far-UV spectrum of the Seyfert 1 galaxy Mrk 509 shows bright
broad \ovi\ $\lambda\lambda$1032,1038 emission as well as
broad \ciii\ $\lambda$977 emission.
Kinematically complex intrinsic absorption in Mrk 509 shows at least 7
distinct components in the Lyman lines, \ovi\ $\lambda\lambda$1032,1038,
and \ciii\ $\lambda$977.
Although many of the \ovi\ and Lyman line components appear to be saturated,
they are not black, implying that partial covering or scattering affects
the absorption.
Only one of the intrinsic absorption components in Mrk 509 is likely to be
associated with the warm X-ray absorbing gas.
Component 5 (near the systemic velocity) has an ionization state and column
density that is in reasonable agreement with the \ovii\ and \oviii\ absorption
edges seen in the ASCA X-ray spectrum.
The high resolution of FUSE and its sensitivity in the \ovi\ band make it
an ideal tool for identifying high-ionization UV absorbers that
may correspond to X-ray warm absorbers.

\acknowledgments

This work is based on data obtained for the Guaranteed Time Team by the
NASA-CNES-CSA FUSE mission operated by the Johns Hopkins University.
Financial support to U. S. participants has been provided by
NASA contract NAS5-32985.
G. Kriss acknowledges additional support from NASA Long Term
Space Astrophysics grant NAGW-4443.

\end{document}